\documentclass{article}
\usepackage{amsmath, amssymb, amsfonts}
\title{ Comments on ''On the Origin of Gravity and the Laws of Newton'', by Erik Verlinde}
\author{Hristu Culetu, \\Ovidius University, Dept.of Physics, \\B-dul Mamaia 124, 900527 Constanta, Romania, \\e-mail : hculetu@yahoo.com}

\begin{document}
\numberwithin{equation}{section}
\pagenumbering{arabic}
\maketitle
\newcommand{\fv}{\boldsymbol{f}}
\newcommand{\tv}{\boldsymbol{t}}
\newcommand{\gv}{\boldsymbol{g}}
\newcommand{\OV}{\boldsymbol{O}}
\newcommand{\wv}{\boldsymbol{w}}
\newcommand{\WV}{\boldsymbol{W}}
\newcommand{\NV}{\boldsymbol{N}}
\newcommand{\hv}{\boldsymbol{h}}
\newcommand{\yv}{\boldsymbol{y}}
\newcommand{\RE}{\textrm{Re}}
\newcommand{\IM}{\textrm{Im}}
\newcommand{\rot}{\textrm{rot}}
\newcommand{\dv}{\boldsymbol{d}}
\newcommand{\grad}{\textrm{grad}}
\newcommand{\Tr}{\textrm{Tr}}
\newcommand{\ua}{\uparrow}
\newcommand{\da}{\downarrow}
\newcommand{\ct}{\textrm{const}}
\newcommand{\xv}{\boldsymbol{x}}
\newcommand{\mv}{\boldsymbol{m}}
\newcommand{\rv}{\boldsymbol{r}}
\newcommand{\kv}{\boldsymbol{k}}
\newcommand{\VE}{\boldsymbol{V}}
\newcommand{\sv}{\boldsymbol{s}}
\newcommand{\RV}{\boldsymbol{R}}
\newcommand{\pv}{\boldsymbol{p}}
\newcommand{\PV}{\boldsymbol{P}}
\newcommand{\EV}{\boldsymbol{E}}
\newcommand{\DV}{\boldsymbol{D}}
\newcommand{\BV}{\boldsymbol{B}}
\newcommand{\HV}{\boldsymbol{H}}
\newcommand{\MV}{\boldsymbol{M}}
\newcommand{\be}{\begin{equation}}
\newcommand{\ee}{\end{equation}}
\newcommand{\ba}{\begin{eqnarray}}
\newcommand{\ea}{\end{eqnarray}}
\newcommand{\bq}{\begin{eqnarray*}}
\newcommand{\eq}{\end{eqnarray*}}
\newcommand{\pa}{\partial}
\newcommand{\f}{\frac}
\newcommand{\FV}{\boldsymbol{F}}
\newcommand{\ve}{\boldsymbol{v}}
\newcommand{\AV}{\boldsymbol{A}}
\newcommand{\jv}{\boldsymbol{j}}
\newcommand{\LV}{\boldsymbol{L}}
\newcommand{\SV}{\boldsymbol{S}}
\newcommand{\av}{\boldsymbol{a}}
\newcommand{\qv}{\boldsymbol{q}}
\newcommand{\QV}{\boldsymbol{Q}}
\newcommand{\ev}{\boldsymbol{e}}
\newcommand{\uv}{\boldsymbol{u}}
\newcommand{\KV}{\boldsymbol{K}}
\newcommand{\ro}{\boldsymbol{\rho}}
\newcommand{\si}{\boldsymbol{\sigma}}
\newcommand{\thv}{\boldsymbol{\theta}}
\newcommand{\bv}{\boldsymbol{b}}
\newcommand{\JV}{\boldsymbol{J}}
\newcommand{\nv}{\boldsymbol{n}}
\newcommand{\lv}{\boldsymbol{l}}
\newcommand{\om}{\boldsymbol{\omega}}
\newcommand{\Om}{\boldsymbol{\Omega}}
\newcommand{\Piv}{\boldsymbol{\Pi}}
\newcommand{\UV}{\boldsymbol{U}}
\newcommand{\iv}{\boldsymbol{i}}
\newcommand{\nuv}{\boldsymbol{\nu}}
\newcommand{\muv}{\boldsymbol{\mu}}
\newcommand{\lm}{\boldsymbol{\lambda}}
\newcommand{\Lm}{\boldsymbol{\Lambda}}
\newcommand{\opsi}{\overline{\psi}}
\renewcommand{\tan}{\textrm{tg}}
\renewcommand{\cot}{\textrm{ctg}}
\renewcommand{\sinh}{\textrm{sh}}
\renewcommand{\cosh}{\textrm{ch}}
\renewcommand{\tanh}{\textrm{th}}
\renewcommand{\coth}{\textrm{cth}}

\begin{abstract}

 We argue that the relativistic Unruh temperature cannot be associated with the bits on the screen, in the form considered by Verlinde. The acceleration $a$ is a scalar quantity (the modulus of the acceleration four vecor) and not a vector.
 
 When the mass $m$ approaches the holographic screen, viewed as a stretched horizon, the shift $\Delta x$ from Verlinde's Eq. (3.15) becomes $c^{2}/a$ and the entropy variation equals $(1/2) k_{B} \Delta N$ , in accordance with Gao's calculations.
 
 Using the Heisenberg Principle we show that the energy on the causal horizon (viewed as a holographic screen) of an inertial observer is proportional to its radius , as for a black hole.

 
\end{abstract}

\textbf{1. Introduction}

 To have a complete theory of quantum gravity we must clarify whether the gravitational interaction is fundamental. One of the remarkable steps toward the nature of gravity is given by black hole thermodynamics and especially the proportionality between entropy and horizon area.
 
 Once Jacobson \cite{TJ} derived Einstein's equations using thermodynamic arguments and the Raychaudhuri equation, Padmanabhan \cite{TP} stressed the emergent character of gravity using the law of equipartition in the horizon degrees of freedom perceived by local Rindler observers.
 
 Recently Verlinde \cite{EV} brought evidences for the entropic nature of the gravitational force. A plethora of papers related to Verlinde's idea emerged in the last weeks. Li and Wang \cite{LW} showed that an UV/IR relation can be derived from the entropic force formalism while Wei et al. \cite{WLW} derived the modified Friedmann equations from the entropic force and the first law of thermodynamics. Other applications related to Verlinde's paper were developed in \cite{TW} \cite{HC1} \cite{LWW} \cite{VS} \cite{YM} \cite{RK}. \\

 \textbf{2. What is $\Delta x$ ?}

 According to Verlinde, $\Delta x$ is a ''displacement'' ( p. 7 of [3]), above Eq. (3.6)). According to the Fig. 2, $\Delta x$ seems to be the distance between the test particle of mass $m$ and the (planar) holographic screen. In Fig. 3
 it is not clear what is $\Delta x $ since the mass $m$ $\textit{is}$ near the spherical screen. Verlinde obtained correctly the law of gravitation (3.13) because the l.h.s. and the r.h.s. of his equation (3.7) are proportional to $\Delta x$. However, that is not a reason not to explain what is exactly $\Delta x$. 
 
 A possible explanation results from the bottom of p. 10, where Verlinde moves the screen instead of the mass $m$. He found correctly that force will not change even though the number of bits on the screen increases . In fact, this is a direct consequence of the Birkhoff theorem stating that the radial pulsations of the holographic screen (or of the mass $M$, which is equivalent to the energy distributed on the screen) will not change the outer gravitational field (and therefore the force on $m$) as long as the spherical symmetry is preserved. 
 
 The fact that the value of $\Delta x$ plays no role for the entropic force has also been observed by Gao \cite{SG} (see also his footnote no. 4, p. 3). He also gave the correct expression for $\Delta S$ in terms of $R$ and $R_{0}$, as it was suggested by Verlinde at p. 10. 
 
 From Eqs. (3.6) and (3.7) it is clear that $x = const.$ leads to $\Delta S = 0$ , whence $F = 0$ , i.e. the entropic force is vanishing. The Newtonian force acting on $m$ is, of course, nonzero when $m$ stays somewhere at a fixed distance from the screen. As Gao \cite{SG} has noticed, Verlinde's causal chain $\Delta x \rightarrow \Delta S \rightarrow F$ does not work and, therefore, gravity is not an entropic force. We agree with Gao's point of view in this respect. However, we consider the implication $\Delta x \rightarrow \Delta S$ to be valid. 
 
 Let us apply this implication for a simple physical situation : take a rocket at rest w.r.t. some inertial reference system I. Let the rocket begins moving with constant acceleration (hyperbolically) w.r.t. I, on a distance, say, $\Delta x$. We know from Special Relativity that a Rindler horizon forms, from the point of view of a traveler inside the rocket. Taking the horizon as the holographic screen, we may say that a change of entropy on the screen will appear because of $\Delta x$. We already have here a natural temperature : the Unruh temperature $T_{U}$ of the thermal bath around the hyperbolic observer, comoving with the rocket. With the help of the Verlinde relations (3.6), (3.7) and the expression of $T_{U}$, one obtains $F = m a$, $m$ being the traveler mass. Here $F$ is the force of inertia exerted on the traveler ( or the force which stretches a spring fixed in the rocket), as if it were atracted by the holographic screen. An entropy increase on the screen leads to an energy (heat) increase \cite{SG} . Where does the energy come from? It comes from the rocket engine. The work done by it equals $T_{U} \Delta S$. The above example could explain the origin of inertial forces : they arise from the entropy increase on the holographic screen (here the Rindler horizon, see \cite{HC2} \cite{AG}). A similar idea has been recently expressed by Lee \cite {JL}, who suggests the inertia is related to dragging the Rindler horizons. \\

 \textbf{3. The time dependent entropy}

   There is another puzzle related to $\Delta x$. In his equation (5.29) Verlinde has written a covariant expression for $\Delta S$ (taking $a$ and $b$ to run from 0 to 3). Suppose $N_{a}$ has a nonzero temporal component ($N_{t} < 0$). We have in this case
\begin{equation}
\frac{\partial S}{\partial t} = - 2\pi k_{B} \frac{mc^{2}}{\hbar} N_{t} 
\label{1}
\end{equation}
If the mass $m$ is at rest w.r.t. the screen ( $\Delta x = 0$), $S$ depends only on time and (1) can be written as 
\begin{equation}
\Delta S = 2 \pi k_{B} \frac{mc^{2}}{\hbar} N_{t} \Delta t
\label{2}
\end{equation}
(the minus sign from (5.29) is not necessary here). 

Eq. (2) shows that we have an entropy change simply because time flows. That is in accordance with Gao's remark that the condition $\Delta x \neq 0$ is not mandatory to get a nonzero gravitational force due to the mass $M$. If the entropy increases on the holographic screens as time proceeds, that means Nature does work for that and the corresponding energy is recovered on the screens. 

We could show the above argument works without any test particle $m$. Let us consider an observer at rest w.r.t. an inertial system in flat space. After a time $\Delta t$, the causal horizon of him expands , covering a sphere of radius $c \Delta t$. Because of the new informations acquired by our observer \cite{HC3}, an entropy variation $\Delta S$ given by 
\begin{equation}
\Delta S = \frac{1}{2} k_{B}~ \Delta N 
\label{3}
\end{equation}
will appear, localized on the causal horizon, considered as a holographic screen. $\Delta N$ above is given by
\begin{equation}
\Delta N = \alpha ~\frac{A}{l_{P}^{2}}
\label{4}
\end{equation}
where $\alpha$ is a constant of the order of unity, $A = 4 \pi (c \Delta t)^{2}$ is the area of the causal horizon after $\Delta t$ and $l_{P} = (G \hbar/c^{3})^{1/2}$ . $\Delta S$ from (3) leads to an energy variation on the screen, given by $\Delta E = T \Delta S$. To get the temperature $T$ we make use of the Heisenberg Principle , applied for the energy per degree of freedom $\epsilon \equiv (1/2) k_{B} T$ 
\begin{equation}
\epsilon ~\Delta t = \beta \hbar
\label{5}
\end{equation}
where $\beta$ is a constant of the order of unity. With $T$ from (5) the energy change becomes
\begin{equation}
\Delta E = 4 \pi \alpha \beta~ \frac{c^{4}}{G}~ c~ \Delta t
\label{6}
\end{equation}
i.e $\Delta E$ is proportional to the radius $c \Delta t$ of the sphere (or to the time elapsed from an arbitrary origin). This resembles the dependence of the black hole mass on the horizon radius : $M_{bh} = (c^{2}/2G) R_{H}$. Therefore, we choose $\alpha = 1/4$ and $\beta = 1/2 \pi$. \\
Dividing Eq. (6) by $\Delta r = c~ \Delta t$, one obtains $\Delta E/\Delta r \equiv F =  c^{4}/2G$, a value akin with that obtained by Easson et al. \cite{EFS} in their study on a cosmological entropic force.

 We could formally define a ''surface gravity'' $\kappa$ on the screen, by analogy with the black hole case
 \begin{equation}
 \kappa = \frac{c^{4}}{4 G M} = \frac{c}{2 \Delta t}
 \label{7}
 \end{equation}
 It is worth to mention that the observer appears to be inside the holographic screen which, in addition, is going away with the velocity of light (see also \cite{HC4} for a model for the black hole interior). A possible explanation of the nature of $\Delta E$ , present even in Minkowski spacetime, has been given in \cite{HC5}.
  We see that all the above physical quantities are time dependent (we should have used, for example, $\Delta T$ instead of $T$). Because of the simplicity of the relations, we think the model works as if we had thermodynamics at equilibrium. For instance, why the black hole temperature is not time dependent?  Since the event horizon (which acts as the causal horizon) is not expanding (the light emitted from the horizon cannot escape outside). Therefore, the Hawking temperature could be obtained from (5) replacing $c \Delta t$ by $4 R_{H}$. In other words, $T$ is constant when the causal horizon (a null surface) is not expanding.\\

 \textbf{4. Acceleration and the entropy gradient}

 There is another thing in Verlinde's paper which deserves to be pointed out. It concerns his Eq. (3.15) . Verlinde considers the particle with mass $m$ approaches the screen , when it should merge with the degree of freedom on the screen. The number of bits $\Delta N$ carried by the particle follows from
 \begin{equation}
 mc^{2} = \frac{1}{2} k_{B} T \Delta N,
 \label{8}
 \end{equation}
  whence Verlinde immediately obtained
\begin{equation}
\Delta S = \frac{1}{2} k_{B} \frac{a \Delta x}{c^{2}} \Delta N.
\label{9}
\end{equation}
What is $\Delta x$ here? Since the particle merged with the microscopic degrees of freedom on the screen, we cannot have here an arbitrary $\Delta x$. To arrive at (3.15) Verlinde used the Unruh formula for the temperature $T$. But Unruh's thermal bath comes from the fact that the hyperbolic (uniformly accelerated) observer has a horizon (the screen plays the role of a local Rindler horizon). Therefore, $\Delta x$ should be $c^{2}/a$, the distance to the horizon (the special role played by this value was also remarked by Lee \cite{JL}) . Hence, Eq. (9) yields
\begin{equation}
\Delta S = \frac{1}{2} k_{B} \Delta N ,
\label{10}
\end{equation}
which is in accordance with Gao's estimation (Ref.[11], p.4)
\begin{equation}
mc^{2} = \Delta E = T \Delta S = \frac{1}{2} k_{B} T \Delta N .
\label{11}
\end{equation}
 Another comment concerns Verlinde's statement ''it may appear somewhat counter-intuitive that the temperature $T$ is related to the vector quantity $a$...''. But in the Unruh expression for $T$, $a$ is not a vector but the modulus of the acceleration 4-vector $a^{b}$. The Unruh formula is purely relativistic and has no a classical counterpart. 
 
 Take, for example, the Rindler metric
 \begin{equation}
 ds^{2} = - c^{2}(1-gx/c^{2})^{2} dt^{2} + dx^{2} + dy^{2} + dz^{2} ,
 \label{12}
 \end{equation}
 where $g$ is the rest-system acceleration and the horizon is located at $x = c^{2}/g$. \\
 An $x = const.$ (static) observer will have the 4 - acceleration
 \begin{equation}
 a^{b} = (0, \frac{g}{1 - gx/c^{2}}, 0, 0) ,
 \label{13}
 \end{equation}
 with the modulus $a = (a^{b} a_{b})^{1/2} = g/(1 - gx/c^{2})$. It equals $a^{x}$ because $a^{b}$ has only one nonzero component (in fact, $a$ is the acceleration appearing in the equation for the hyperbolic trajectory, $x^{2} - c^{2}t^{2} = (c^{2}/a)^{2})$. 
 
 We see that $a$ is $x$ - dependent and is equal to $g$ at the origin of coordinates (i.e., far from the horizon). The surface gravity on the horizon can be obtained from
 \begin{equation}
 \kappa = \sqrt{a^{b} a_{b}} ~\sqrt{- g_{tt}} |_{x = c^{2}/g} = g 
 \label{14}
 \end{equation}
 and, as in the case of the surface gravity ($ = c^{4}/4GM$ ) for a Schwarzschild black hole, it is measured from infinity, which is equivalent to ''far from the horizon''. \\
 Therefore, the use of $\ddot{x}$ or $a^{x}$ in the Unruh formula is not appropriate. A similar opinion has recently been expressed by Cai et al. \cite{CCO}. \\

 \textbf{5. Conclusions}\\
 We pointed out in this letter few comments upon Verlinde's paper on the origin of gravity. Since the gravitational force is nonzero even when Verlinde's $\Delta x$ is vanishing, we reached at the conclusion that gravity cannot be an entropic force in the form formulated by the author.
 
 From the covariant form of the entropy change, we conjectured that a time dependent entropy on the holographic screen (viewed as the Rindler horizon) is at the origin of inertial forces.


\begin{thebibliography} {20}
\bibitem {TJ}
T. Jacobson, Phys. Rev. Lett. 75, 1260 (1995).
\bibitem{TP}
T. Padmanabhan, ArXiv : 0912.3165 [gr-qc].
\bibitem{EV}
E. Verlinde, ArXiv : 1001.0785 [hep-th].
\bibitem{LW}
M. Li and Y. Wang, ArXiv : 1001.4466 [hep-th].
\bibitem{WLW}
S. Wei, Y. Liu and Y. Wang, ArXiv : 1001.5238 [hep-th].
\bibitem{TW}
Y. Tian and X. Wu, ArXiv : 1002.1275 [hep-th].
\bibitem{HC1}
H. Culetu, ArXiv : 1001.4740 [hep-th].
\bibitem{LWW}
Y. Liu, Y. Wang and S. Wei, ArXiv : 1002.1062 [hep-th].
\bibitem{VS}
I. Vancea and M. Santos, ArXiv : 1002.2454 [hep-th].
\bibitem{YM}
Y. Myung, ArXiv : 1002.0871 [hep-th].
\bibitem{RK}
R. A. Konoplya, ArXiv : 1002.2818 [hep-th].
\bibitem{SG}
S. Gao, ArXiv : 1002.2668 [gr-qc].
\bibitem{JL}
J.-W. Lee, ArXiv : 1003.4464 [hep-th].
\bibitem{HC2}
H. Culetu, Int. J. Mod. Phys.D15, 2177 (2006).
\bibitem{AG}
F. Alexander and U. Gerlach, Phys. Rev. D44, 3887 (1991) (ArXiv : gr-qc/9910086).
\bibitem{HC3}
H. Culetu, Found. Phys. 35, 1511 (2005).
\bibitem{HC4}
H. Culetu, Int. J. Mod. Phys. A24, 1593 (2009).
\bibitem{HC5}
H. Culetu, ArXiv : 0905.3474 [hep-th].
\bibitem{EFS}
D. A. Easson, P. H. Frampton and G. F. Smoot, ArXiv ; 1002.4278 [hep-th].
\bibitem{CCO}
R. Cai, L. Cao and N. Ohta, ArXiv : 1001.3470v2 [hep-th].


\end{thebibliography}
\end{document}